# Magnetic-field dependence of valley splitting for Si quantum wells grown on tilted SiGe substrates

Seungwon Lee and Paul von Allmen

*Jet Propulsion Laboratory, California Institute of Technology, Pasadena, CA 91109*

(Dated: July 15, 2006)

**Abstract**

The valley splitting of the first few Landau levels is calculated as a function of the magnetic field for electrons confined in a strained silicon quantum well grown on a tilted SiGe substrate, using a parameterized tight-binding method. For a zero substrate tilt angle, the valley splitting slightly decreases with increasing magnetic field. In contrast, the valley splitting for a finite substrate tilt angle exhibits a strong and non-monotonous dependence on the magnetic field strength. The valley splitting of the first Landau level shows an exponential increase followed by a slow saturation as the magnetic field strength increases. The valley splitting of the second and third Landau levels shows an oscillatory behavior. The non-monotonous dependence is explained by the phase variation of the Landau level wave function along the washboard-like interface between the tilted quantum well and the buffer material. The phase variation is the direct consequence of the misorientation between the crystal axis and the confinement direction of the quantum well. This result suggests that the magnitude of the valley splitting can be tuned by controlling the Landau-level filling factor through the magnetic field and the doping concentration.

Experimental investigations using a variety of measurement techniques have shown that the valley splitting for a two-dimensional electron gas confined in either Si metal oxide semiconductor field effect transistors (MOSFET) or modulation doped Si/SiGe heterostructures strongly depends on the magnetic field strength and on the Landau level filling factor [1]. This strong dependence has been attributed to various mechanisms including i) the misorientation of the growth direction [2], ii) electron exchange interaction [3], iii) electric breakthrough mechanisms [4], and iv) surface scattering [5]. The valley-splitting in the presence of magnetic field has been widely studied using effective-mass approximations, yet no satisfactory agreement has been achieved with experimental data. For example, the linear magnetic field dependence of the valley splitting for the first Landau level has not yet been obtained in simulations. Calculations using the effective-mass approximation rely on first order perturbation and an ad hoc interface potential to include the valley splitting. Since experimental and numerical data are not in agreement, calculations that do not require perturbation theory nor empirical parameters to describe the valley splitting are desirable. Valley splitting for silicon quantum wells (QW) has been calculated previously using a parameterized tight-binding method [6] and the effects of a misoriented substrates has been reported using the same approach [7]. However, to our knowledge no study of the magnetic field dependence using an atomic level approach has been attempted so far.

The purpose of the present paper is to report valley-splitting calculations in presence of a magnetic field for a strained Si QW grown on a tilted substrate using a parameterized tight-binding model, where the valley splitting is obtained by directly diagonalizing the Hamiltonian without resorting to any perturbation method. In particular, the valley splitting in the single-particle picture arising from the misorientation between the crystal axis and the confinement axis is examined. These calculations broadly confirm the effective mass calculations for the first Landau level and provide an intuitive understanding of the non-monotonous magnetic field dependence for the second and third Landau levels.

For this study, the tight-binding parameters for silicon are taken from reference 8. The strain effect on the Hamiltonian is incorporated through off-diagonal elements [8]. The magnetic field effect is described by applying a gauge-invariant Peierls substitution to the off-diagonal elements [9]. The Peierls substitution incorporates the vector potential into the tight-binding Hamiltonian without introducing additional fitting parameters. The confinement potential provided by the buffer material is approximated through passivating dangling bonds at the interface [10]. This approximation does not alter the qualitative behavior of the valley splitting while it reduces the valley splitting magnitude due to the weakening of the wave function amplitude at the interface.

The atomic structure of the strained silicon quantum well grown on a tilted substrate forms monolayer or bilayer steps along a preferential direction with irregular step periods [11]. As an initial study, we model the atomic structure with regular monolayer steps aligned along the crystal axis $x$ [100] (see Fig. 1). The separation between the steps ($L_s$) is related to the tilt angle of the substrate, $\theta$ by $\tan\theta = h/L_s$ where $h$ is the monolayer step height. The strain in the quantum well due to the lattice mismatch between the Si and $Si_{1-x}Ge_x$ materials is modeled with a uniform biaxial strain with strain tensor values $\varepsilon_{xx} = \varepsilon_{yy} = 0.0125$ and $\varepsilon_{zz} = -0.0103$, which correspond to a Ge concentration $x = 0.3$. The dimensions of the modeled structure are specified in Figure 1.

For a magnetic field $B$ applied along the confinement direction $z'$, we choose a vector potential $A = (0, By, 0)$. This asymmetric gauge breaks the periodicity of the Hamiltonian along *x'*, but maintains the periodicity along the crystal axis y [010]. In order to ensure that the finite length of the modeled system does not affect the confinement of the Landau level, the length along *x'* ($L_{x'}$) is chosen to be at least nine times as big as the effective confinement length of the Landau level.

Figure 2 shows the first three Landau level energies calculated as a function of the applied magnetic field strength. The wave function along x' plotted in the inset shows that the size of the modeled system is sufficiently large to accommodate the confinement of the Landau level and at the same time to avoid any spurious effect due to the arbitrary finite-length confinement. The slope of the magnetic field dependence suggests that the effective mass of the conduction-band electron along x' is about 0.24 $m_0$, where $m_0$ is the free electron mass. This value agrees with that obtained from rotating the effective mass tensor by the tilt angle θ (3.5 degrees in the modeled system). The tight-binding model yields the effective masses $m_{xx}$=0.24 $m_0$ and $m_{zz}$=0.74 $m_0$ for the strained silicon. By rotating the effective mass tensor, the effective mass along x' becomes 0.23 $m_0$ given by $m_{x'x'}^{-1} = m_{xx}^{-1}\cos^2\theta + m_{zz}^{-1}\sin^2\theta$, in agreement with the Landau level variation.

Figure 3(a) shows the valley splitting for the first three Landau levels as a function of the magnetic field strength. The valley splitting of the first level shows an exponential increase followed by saturation as the magnetic field increases, while that of the second and third levels shows an oscillatory behavior. In order to examine whether the misorientation between the crystal axis (z) and the confinement direction (z') is responsible for this strong magnetic-field dependence, the valley splitting in the absence of misorientation is also calculated. Figure 3(b) shows that the strong magnetic field dependence vanishes when the quantum well is grown on a substrate with zero tilt angle. The valley splitting slightly decreases from 7.15 meV to 6.95 meV as the magnetic field increases from 0 T to 30 T. This result shows that the strong dependence of the valley splitting on the magnetic field and the Landau level index is the direct consequence of the misorientation between the crystal axis *z* and the confinement direction *z'*.

The comparison between Figure 3(a) and (b) illustrates another misorientation effect, which is a reduction in the magnitude of the valley splitting. In the presence of misorientation, the valley splitting reaches at most 5% of the valley splitting with zero tilt angle. In addition, a previous tight-binding study including a band structure calculation shows that the valley splitting completely vanishes at zero magnetic field with a finite tilt angle [7].

The strong magnetic-field dependence of the valley splitting can be interpreted explicitly in terms of simple arguments using the effective mass approximation. A similar interpretation has been reported for the valley splitting of the first Landau level in Ref. [12]. Wave functions for the two lowest Z-valley states confined in a QW of width *W*, with an infinite potential barrier, and with a magnetic field B applied along the growth direction z' can be approximated by the following expression:

$$\psi_n^{\pm}(x',z') = \cos(\pi z'/W) H_n(x'/\ell_B) e^{-x'^2/2\ell_B^2} \, e^{\mp i k_0 \sin\theta x' \pm i k_0 \cos\theta z'}, \qquad (1)$$

where $H_n(x)$ is a Hermite polynomial and $\ell_B$ is the Larmor radius given by $\sqrt{eB/\hbar}$. Note that the normalization constant for the wave function is dropped for simplicity. The phase factors in the x' and z' directions are $k_0 \sin\theta \, x'$ and $k_0 \cos\theta \, z'$, respectively, because the conduction band minima in the rotated coordinate frame *(x'y'z')* are at $k = \mp k_0 \sin\theta \, \hat{e}_{x'} \pm k_0 \cos\theta \, \hat{e}_{z'}$, where $k_0 \cong 0.15 \, (2\pi/a)$ is the position of the conduction band minimum in the folded Brillouin zone for bulk silicon.

The two degenerate valley wave functions $\psi_n^{\pm}(x',z')$ are coupled due to a translational-symmetry-breaking potential at the interfaces located at $z' = W/2$, $z' = -W/2$,. We approximate the symmetry breaking potential as $V(r) = v_1 g(z'-W/2) + v_2 g(z'+W/2)$, where $g(z')$ is a highly localized function centered at zero with its localization range being about one atomic monolayer. The exact form of $g(z')$ is not critical for this argument. The constants $v_1$ and $v_2$ are determined by the atomic and bonding environment at the interface. The coupling strength between $\psi_n^{\pm}(x',z')$ through the symmetry breaking potential determines the valley splitting:

$$\Delta E_n = 2\left|\langle \psi_n^+ a | V(r) | \psi_n^- \rangle\right|$$
$$= \lambda \int dx' H_n^2\left(\sqrt{\frac{eB}{\hbar}}x'\right) e^{-\frac{eB}{\hbar}x'^2} e^{i2k_0 \sin\theta x'} \quad (2)$$

where the constant $\lambda$ results from the integration over z'.

The resulting expression contains the phase variation $2k_0 \sin\theta x'$ in the integrant, which leads to the rich magnetic field dependence of the valley splitting. When the magnetic field is zero, the wave function extends over an infinite range along x', and the phase variation therefore leads to a complete cancellation of the valley coupling in agreement with previous band-structure calculations [7]. As the magnetic field increases, the extent of the wave function along x' decreases and the cancellation effect of the phase variation is incomplete. In particular, the relative peak locations of the second and third Landau level wave functions affect the phase interference between the peaks, resulting in an oscillatory behavior of the valley splitting. The same equation for the valley splitting also explains the disappearance of the strong dependence at zero tilt angle. Without the phase variation, the integration over x' leads to a constant independent of the magnetic field and Landau level index. This prediction is consistent with the weak dependence obtained by the direct tight-binding calculation shown in Figure 3(b).

Equation (2) provides an insight into the interplay between the two characteristic lengths that determine the magnitude of the valley splitting. The Larmor radius $\ell_B$ characterizes the extent and the peak separation in the Landau level wave function along x', and the step separation ($L_s \sim \tan\theta$) represents the period of the phase variation along x'. When the valley splitting equation is expressed in terms of the ratio between these two length scales, it provides a universal dependence curve with regard to the tilt angle variation because the interplay between the two length scales remains the same. In Figure 3(c) and 3(d), the valley splitting dependence on the magnetic field is plotted with respect to the ratio of the Larmor radius to the step separation for a tilt angle of 3.5 degrees and 7.1 degrees, respectively. The plots illustrate that irrespective of the tilt angle, the valley splitting for the second and third Landau levels vanishes at $\ell_B/L_s \approx 2.5$ and 4.8, respectively, where the wave function peaks are located such that the phase interference is destructive.

    The strong dependence of the valley splitting on the Landau level index and the magnetic field offers an opportunity to engineer the magnitude of the valley splitting for device applications. A large valley splitting can be achieved by tuning the Landau-level filling factor through changes in the magnetic field or doping concentration so that the Landau-level at the Fermi level has a high Landau index and the multiple peaks of the wave function are located relatively to each other such that the phase interference between the peaks becomes constructive.

    We make a few comparative remarks between these calculation results and relevant experiments. With regard to the characteristics of the magnetic-field dependence, the calculated result is in agreement with experimental results although the exact dependence is different [1]. The experiments give a linear dependence of the valley splitting for the first Landau level. This discrepancy between theory and experiment suggests that the valley splitting arising from the misorientation effect in the single-particle picture does not account for the whole magnitude of

the valley splitting. Other mechanisms such as many-body interactions may be responsible for the enhanced valley splitting at low magnetic fields as suggested in prior calculations [3, 13].

As for the valley-splitting dependence on the Landau level index, the calculation is consistent with the recent magneto-transport measurements by Lei et al [1]. The experiment demonstrates that the valley splitting from different sides of the coincidence region differs by a factor of 3. The valley splitting from each side of the coincidence arises from a different Landau level. Our calculations show that due to the different magnetic-field dependence, the valley splitting of a different Landau level can be significantly different for some range of magnetic fields.

Recently, a remarkable distinct behavior of the valley splitting was observed for electrons confined in $SiO_2/Si/SiO_2$ quantum wells on silicon-on-insulator structures [13]. The valley splitting does not change with increasing magnetic field, and is strongly asymmetric with respect to the electrical gate bias, indicating that topological differences (atomic terracing disorder) between the two $SiO_2/Si$ interfaces (thermal-oxide/Si and buried-oxide/Si) lead to the asymmetric valley splitting. According to the results presented in this paper, these new observations can be understood by the competition between disorder effect and misorientation effect. The highly disordered interface between the buried oxide and Si overshadows the misorientation effect, and thus enhances the valley splitting and removes its magnetic-field dependence.

In summary, the valley splitting of the first few Landau levels for a strained silicon QW grown on an unstrained tilted SiGe substrate is calculated using a tight-binding model. Specifically, the valley splitting resulting from the misorientation effect in the single particle picture is investigated. A strong magnetic field dependence of the valley splitting and Landau level index is observed and is attributed to the phase variation of the wave function along the washboard-like interface between the quantum well and the buffer. The phase variation arises from the misorientation between the crystal axis and the growth direction of the silicon QW grown at a tilted angle. The strong dependence can be exploited to engineer the magnitude of the valley splitting.

This work was performed at the Jet Propulsion Laboratory, California Institute of Technology under a contract with the National Aeronautics and Space Administration. Funding was provided under a grant from National Security Agency.

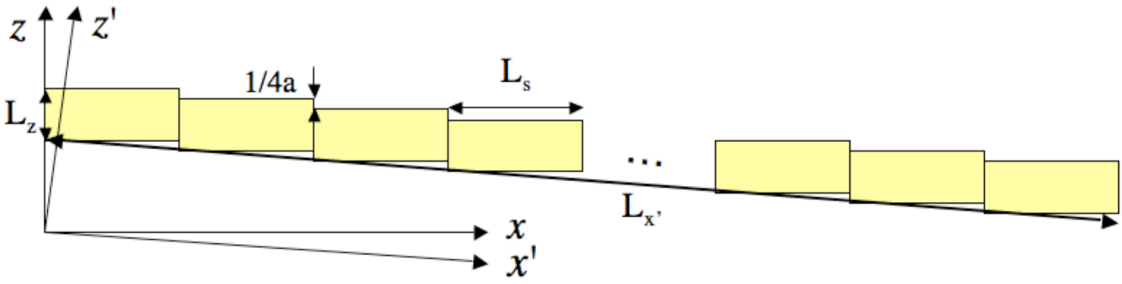

**Figure 1:** Geometry of a quantum well grown on a tilted substrate. The crystal symmetry directions are along *x* [100] and *z* [001]. The QW confinement direction is along *z'*, and atomic steps are formed along *x'*. The step height is one atomic layer (*a/4*). In the modeled system, the step separation $L_s$ is *4a* and the quantum well width $L_z$ is *5a*. This structure corresponds to a tilt angle of 3.5 degrees.

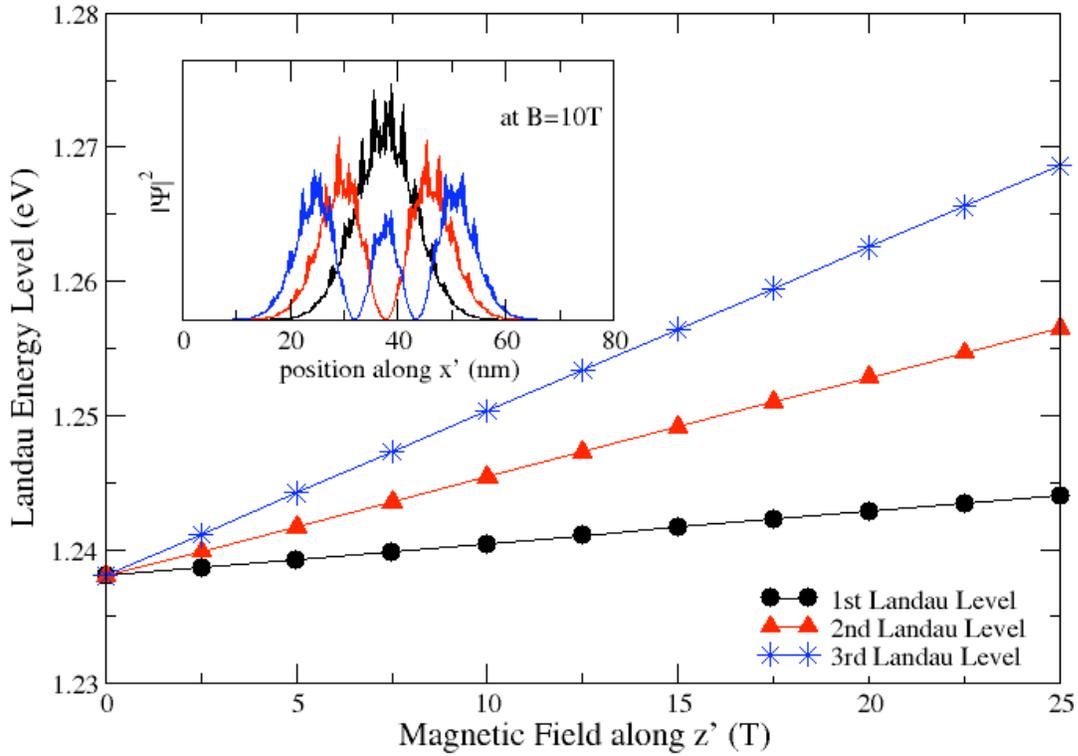

**Figure 2:** First three Landau level energies as a function of magnetic field strength for a 5a wide strained silicon QW with a tilt angle of 3.5 degrees when the magnetic field is applied along the confinement direction z'. The inset shows the wave function square integrated over y and z', illustrating its confinement along x'. The finite length of the modeled system (80 nm for B=10T) is sufficiently long to exclude artificial finite-length simulation effect.

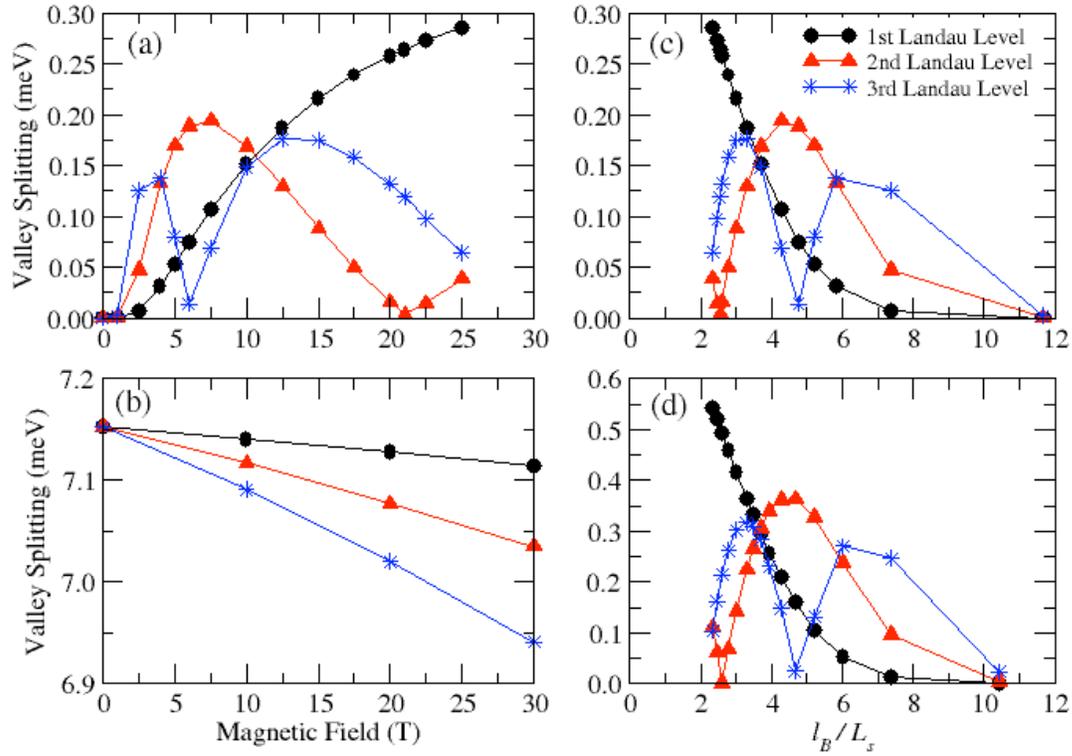

**Figure 3:** Valley splitting of the first three Landau levels as a function of magnetic field strength for a silicon QW (a) with a tilt angle of 3.5 degrees and (b) with a zero tilt angle. The valley splitting as a function of the ratio of the Larmor radius to the step separation for a silicon QW (c) with a tilt angle of 3.5 degrees and (d) with a tilt angle of 7.1 degrees. The QW is 20 atomic layer wide. The magnetic field is applied along the confinement direction, which is z [001] for a no-tilt QW and z' for a tilted QW.